\documentclass[sigconf]{aamas} 

\usepackage{balance} 

\usepackage{bm}
\usepackage[scr=boondoxo]{mathalpha}
\usepackage{xcolor}
\usepackage[ruled, linesnumbered]{algorithm2e}
\usepackage{hyperref}
\usepackage{caption}
\usepackage{subcaption}

\setcopyright{ifaamas}
\acmConference[AAMAS '24]{Proc.\@ of the 23rd International Conference
on Autonomous Agents and Multiagent Systems (AAMAS 2024)}{May 6 -- 10, 2024}
{Auckland, New Zealand}{N.~Alechina, V.~Dignum, M.~Dastani, J.S.~Sichman (eds.)}
\copyrightyear{2024}
\acmYear{2024}
\acmDOI{}
\acmPrice{}
\acmISBN{}

\acmSubmissionID{933}

\title[Holonic Learning]{Holonic Learning: A Flexible Agent-based Distributed Machine Learning Framework}

\author{Ahmad Esmaeili}
\affiliation{
  \institution{Purdue University}
  \city{West Lafayette}
  \country{United States}}
\email{aesmaei@purdue.edu}

\author{Zahra Ghorrati}
\affiliation{
  \institution{Purdue University}
  \city{West Lafayette}
  \country{United States}}
\email{zghorrat@purdue.edu}

\author{Eric T. Matson}
\affiliation{
  \institution{Purdue University}
  \city{West Lafayette}
  \country{United States}}
\email{ematson@purdue.edu}

\begin{abstract}
Ever-increasing ubiquity of data and computational resources in the last decade have propelled a notable transition in the machine learning paradigm towards more distributed approaches. Such a transition seeks to not only tackle the scalability and resource distribution challenges but also to address pressing privacy and security concerns. 
To contribute to the ongoing discourse, this paper introduces Holonic Learning (HoL), a collaborative and privacy-focused learning framework designed for training deep learning models.  
By leveraging holonic concepts, the HoL framework establishes a structured self-similar hierarchy in the learning process, enabling more nuanced control over collaborations through the individual model aggregation approach of each holon, along with their intra-holon commitment and communication patterns. HoL, in its general form, provides extensive design and flexibility potentials. For empirical analysis and to demonstrate its effectiveness, this paper implements HoloAvg, a special variant of HoL that employs weighted averaging for model aggregation across all holons. The convergence of the proposed method is validated through experiments on both IID and Non-IID settings of the standard MNISt dataset. Furthermore, the performance behaviors of HoL are investigated under various holarchical designs and data distribution scenarios. The presented results affirm HoL's prowess in delivering competitive performance particularly, in the context of the Non-IID data distribution. 
 
\end{abstract}


\newcommand{\BibTeX}{\rm B\kern-.05em{\sc i\kern-.025em b}\kern-.08em\TeX}

\begin{document}


\pagestyle{fancy}
\fancyhead{}


\maketitle 


\section{Introduction}
Today's interconnected world, marked by the proliferation of information and computational capabilities across vast digital landscapes, has led to critical challenges concerning the scalability, adaptability, speed, and support for diversity in the traditional centralized learning models. As a solution to such limitations and harnessing the collective power of the networked devices---whether servers or edge devices---recent decade has witnessed a substantial growth in the rise of distributed learning methodologies that not only expedite training process but also enhances robustness and privacy. 

Federated Learning (FL) \cite{mcmahan2017communication} stands out as one of the prominent distributed learning approaches attracting attentions from both academia and industry \cite{yang2019federated}. By empowering end users to participate in the collaborative learning process while retaining ownership of their data, FL ensures both data security and privacy on one hand, and leveraging the potential of computationally constrained devices on the other. Despite its apparent simplicity in its original format, FL introduces a series of intricate open problems and challenges tackled by a plethora of research work. Communication efficiency, convergence rates, privacy preservation, addressing potential adversarial attacks, robustness across heterogeneous devices and data distributions, non-identically and independently distributed (non-IID) data, mitigating potential aggregation bias, and ensuring fairness in model updates are among the nuanced and critical open problems that arise within the realm of Federated Learning \cite{kairouz2021advances}. 

Emerging real-world scenarios and applications have given rise to new trends in methodologies that focus on relaxing some of FL's core assumptions. Some of the noteworthy examples include peer-to-peer Learning \cite{jiang2017collaborative, vanhaesebrouck17a}, edge-based learning \cite{wang2019adaptive,tran2019federated}, and split learning \cite{gupta2018distributed, vepakomma2018split}. In fully decentralized Peer-to-Peer (P2P) learning the objective revolves around replacing the central orchestration server with a peer-to-peer communications topology between the clients to mitigate the risk of single-point of failure and address the potential communication bottlenecks in large-scaled systems. In edge-based FL system, on the other hand, the emphasis lies in relocating the parameter server to the vicinity of edge devices to enhance latency and strike a balance between computation and communication. Unlike the previous approaches, split learning introduces another dimension of parallelism and shifts the focus from data partitioning to splitting the execution of the learning model into layers across multiple clients and the server. 

Hierarchical approaches have consistently been instrumental in shaping the development and structure of intricate systems, and FL is no exception to this pattern. In Hierarchical FL approaches, the clients are clustered into groups to reduce communication overheads and add an extra privacy layer through periodic model updates within and across groups \cite{rana2023hierarchical}. The studies reported in \cite{abad2020hierarchical, liu2020client, xu2021adaptive, hosseinalipour2022multi} exemplify noteworthy contributions that center on enhancing energy consumption, latency, and communication efficiency. In \cite{abad2020hierarchical}, a two-tier hierarchical FL architecture is employed for wireless heterogeneous cellular networks. This structure clusters mobile users into localized cells, leveraging local FL within micro-cell clusters and aggregating model updates at the macro-level through base stations. The HierFAVG model introduced in \cite{liu2020client} presents a three-layer hierarchy---client-edge-cloud---extending FL with a two-level average-based model aggregation approach. In \cite{xu2021adaptive}, a dual-level framework is proposed, comprising mobile devices and edge/cloud servers. This framework optimizes resource allocation and aggregation intervals adaptively. Finally, the work outlined in \cite{hosseinalipour2022multi}, unlike the proceeding models, employs an arbitrary number of levels in its design. Their proposed tree-like structure has a fixed hierarchical topology and is constructed based on the physical proximity and connectivity of devices. Notably, their model utilizes a device-to-device (D2D) consensus method to update the local models at the lowest level, followed by a sequence of recursive aggregations through the upstream path, culminating in a central server.

This paper presents a holonic solution for the distributed learning problem. The multi-level architecture of holonic systems embodies an arrangement of self-contained entities known as holons, forming a self-similar structure rooted in the whole/part relationship. Benefitting from their balance between local autonomy and coordinated decision-making, holonic systems adeptly tackle challenges such as adaptability, fairness, and scalability. Additionally, their decentralized authority and decision-making approach mitigate the risk of a single point of failure, further enhancing their resilience. Holonic solutions have already been employed in the design and management of machine learning systems. As an example, the Hierarchical Agent-based Machine LEarning PlaTform (HAMLET) presented in \cite{esmaeili22hamlet} leverages holons to organize machine learning resources distributed across an array of networked devices and orchestrates automatic parallel training and testing tasks. Our paper, in contrast, primarily focuses on distributed learning and introduces a novel holonic collaborative learning methodology.

\begin{figure}
    \centering
    \includegraphics[width=\columnwidth]{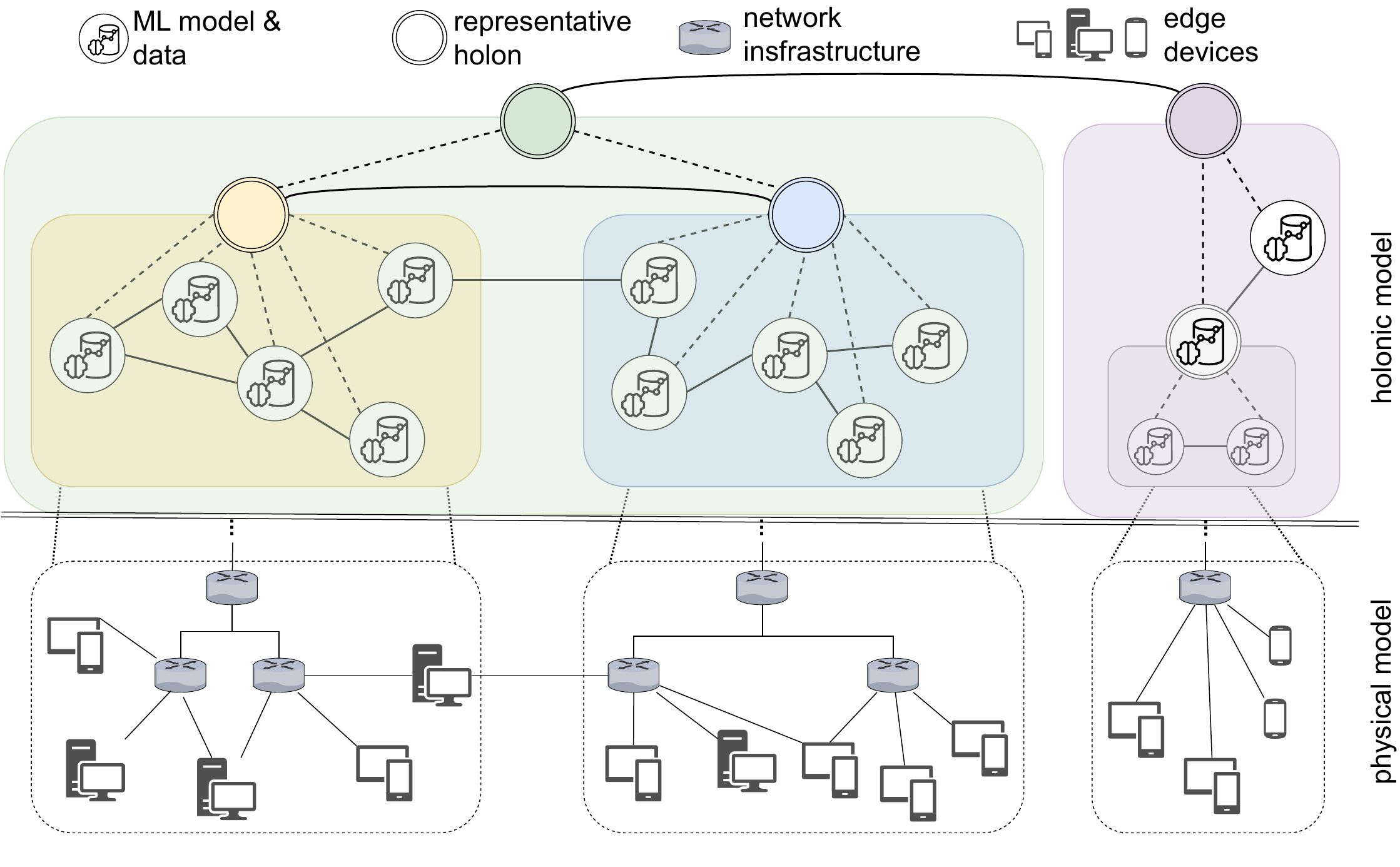}
    \caption{An example HoL model built over an IoT network}
    \label{fig:example_struct}
\end{figure}

Figure \ref{fig:example_struct} portrays an illustrative multi-level holonic model built over a set of connected devices through grouping them together at different levels of granularity. The construction of such structure indeed constitutes a highly challenges phase, potentially contingent upon various factors ranging from physical constraints to trust and security considerations. To maintain our main focus and streamline the analysis, we assume that the holonic construction is preexisting and accessible. We will return to this figure in the subsequent section as we succinctly review the fundamentals of holonic concepts. 

The presented Holonic Learning (HoL) model generalizes the conventional distributed approaches by viewing the holons of each level, as a network of self-contained agents (workers or clients, depending on the application) that collaborate to optimize the objective model parameters regardless of paying attention to the details of learning process that each agent pursues. In other words, each holon member of the level has the autonomy to chose to learn while being committed to the goal of the superordinate it is a member of. This model differs from fully decentralized peer-to-peer learning models with direct inter-agent coordination as it facilitates enforcing diverse set of policies at different levels. On the other hand, thanks to its self-similar structure, it is distinct from hierarchical approaches that require clear lines of authority on one hand and relying on a central server to dictate the next steps. The main contributions of this paper are highlighted as follows:
\begin{itemize}
    \item We present a collaborative learning framework that consolidates existing distributed machine learning solutions. By harnessing both vertical and horizontal interactions, and drawing inspiration from the self-similar concepts of holonic structures, our framework enables the integration of more sophisticated collaboration patterns while maintaining an intuitive design.
    \item Through your empirical analysis, you provide concrete evidence of HoL's effectiveness in achieving consistent convergence behavior across various holarchies and data distribution settings. The comparison of performance metrics, such as training loss and test accuracy, demonstrates the advantages of the holarchical structure in improving learning outcomes compared to plain collaboration networks.   
\end{itemize}

The paper is organized as follows: Section \ref{sec:prelim} reviews the fundamental principles of holonic systems and the holonic multi-agent concepts employed in this study. Section \ref{sec:hlmodel} delves into the formulation of our holonic learning model and presents the details of a more specific use-case analyzed in this paper. Section \ref{sec:experiments} elaborates on the experiments conducted and examines the model's characteristics from multiple perspectives. Finally, Section \ref{sec:conclusion} concludes the paper and offers suggestions for future research directions.


\section{The Preamble}\label{sec:prelim}

\subsection{Holonic Structures}\label{sec:holon}
Formally introduced by Arthur Koestler, the term ``holon'' refers to a philosophical concept describing entities that are simultaneously whole in themselves and parts of larger systems \cite{koestler1968ghost}. In other words, a self-contained holon embodies both individuality and integration, which helps to model complex systems and understand how they evolve in response to changing circumstances. Although it was originally used to describe the self-similar hierarchical structure of biological and social system, it has found its way into many disciplines such as computer science \cite{esmaeili22hamlet}, engineering \cite{benabdellah2021smartdfrelevance}, physics \cite{capra1984turning}, and manufacturing \cite{valckenaers1997holonic}, to name a few. This section highlights the key definitions related to the holonic systems that are heavily used in the subsequent context.

As stated above, the key idea of holons is their dual nature, or as described in \cite{koestler1978janus}, how they behave like a two-faced Janus: the face representing the whole looks inward and the other looks up, or outward. That is at the same time that each holon is independent with its own characteristics and functions, it is also a relatively more concrete component of larger systems. Applying such a whole-part relationship to different entities of a system create a multi-level interconnected structure of holons, where the behavioral variations of each holon at any level is characterized by operational variation of its subordinate holons in a lower level. In the holonic terminology, such a hierarchical architecture is called ``holarchy'', and the higher-level and lower-level holons are respectively referred to as ``super-holons'' and ``sub-holons''. In this paper, we also use the terms ``terminal holons'' and ``non-terminal'' holons to refer to holons at the lowest level of the holarchy and the ones in the intermediate layers, respectively. 

When incorporating holonic principles into the realm of computer science and the design of distributed autonomous systems, the ``Holonic Multiagent Systems (HMAS)'' emerges as an exceptionally promising and broadly applicable model to embrace. As originally proposed in \cite{gerber1999holonic}, a holon (holonic-agent) is formed by a group of agents giving up part of their autonomy to the super-holon and through the designation of one agent as the representative called ``head''. With administrative and/or directive responsibilities, the head plans for the holon on basis of its subordinates. With that said, a holon is formally characterized by the tuple $\left<head, subholons, C\right>$, where $C$ denote commitments, i.e., the itera-holon relationships agreed by all members \cite{fischer2003holonic}. Based on this formalization, each agent $a$ of a MAS constitutes an atomic (terminal) holon $\left<\{a\}, \{a\}, \emptyset\right>$ at the lowest level of the holarchy, and there are exponential ways to form each upper-level holon based on its subordinates.

\begin{figure}
    \centering
    \includegraphics[width=\columnwidth]{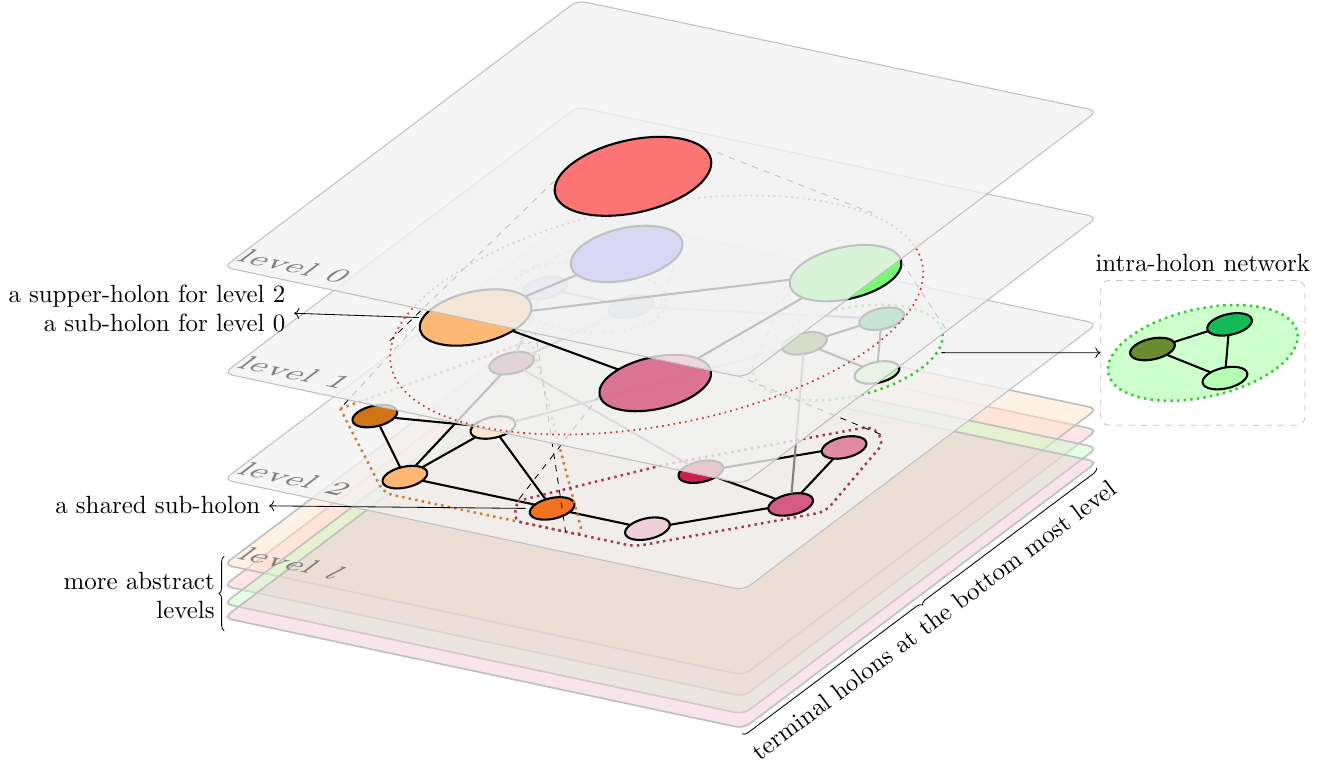}
    \caption{An example holarchical network. The levels are numbered from top to bottom, and different color shades represent the heterogeneity of the holons. }
    \label{fig:holarchy}
\end{figure}

There are multitude of research efforts, spanning both domain-specific and generic contexts, that have delved into the dynamic formation of holarchies, such the ones reported in \cite{esmaeili2017socially, hilaire2008adaptive, barbosa2015self} and \cite{ulieru2002emergence} to name a few. Due to our focus on learning, this paper opts for a predetermined, static structure for the utilized holarchy, while acknowledging that relaxing this assumption could potentially pave the way for novel research trajectories within the realm of distributed learning. To emphasize both inter- and intra-holon interactions, we define a ``holarchical network'' as depicted in Fig. \ref{fig:holarchy}. In constructing a holarchy, super-holons are virtual concepts overseen by head holons. These head holons can be specialized terminal holons designated solely for this role, or they can be existing holons selected (temporarily or permenantly) to serve as heads while concurrently fulfilling their duties as subordinates at higher levels. To simplify the discussion in this paper, we assume the former cases in our HoL framework.

\subsection{Notations}
We use the following notations throughout this paper. Please note that we use the term ``holon'' to refer to the nodes of the HoL model, not only to comply with the holonic terminologies, but also due to the fact that it facilitates referring to the both physical and abstract nodes residing at different level of the proposed HoL model. At the physical level, one can safely assumes each holon represent an edge device.
\begin{itemize}
    \item $\mathscr{h}^{(l)}_{i}$: the $i$-th terminal holon at level $l$. 
    \item $\bm{\mathscr{h}}^{(l)}_{i}$: the $i$-th super-holon at level $l$. This notations is also used to denote the head of the holon.
    \item $\mathsf{h}^{(l)}_{i}$: the $i$-th holon at level $l$ regardless of specifying if it is terminal or non-terminal. 
    \item $n^{(l)}_{i}$: the number of subordinates in the holon $\bm{\mathscr{h}}^{(l)}_{i}$.
    \item $\mathcal{M}^{(l)}_{i}$: the learning model of holon $\mathscr{h}^{(l)}_{i}$.
    \item $\bm{\theta}^{(l)}_{i}(t)$: the model parameters of holon $\mathscr{h}^{(l)}_{i}$ at the end of iteration $t$.
    \item $\mathcal{F}(\bm{\theta}^{(l)}_{i};\bm{x}^{(j)}, y^{(j)})$: the loss function of holon $\mathscr{h}^{(l)}_{i}$ for data sample $(\bm{x}^{(j)}, y^{(j)})$. For the sake of brevity we write it as $\mathcal{F}^{(l)}_{i}(\bm{\theta};\bm{\xi}^{(j)})$ or $\mathcal{F}^{(l)}_{i}(\bm{w})$ depending on the detail level.
    \item $\mathbb{X}^{(l)}_{i}$: the private dataset used for training holon $\mathscr{h}^{(l)}_{i}$'s model.
     \item ${C}^{(l)}_{i}$: the local update stopping criteria of holon $\mathscr{h}^{(l)}_{i}$.
    \item $\mathcal{G}^{(l)}_{i}$: the communication graph of super-holon $\bm{\mathscr{h}}^{(l)}_{i}$.
    \item $\text{Sup}(\mathsf{h}^{(l)}_{i})$: the superior(s) of holon $\mathsf{h}^{(l)}_{i}\in\{\mathscr{h}^{(l)}_{i}, \bm{\mathscr{h}}^{(l)}_{i}\}$ in the holarchy.
    \item $\text{Sub}(\mathsf{h}^{(l)}_{i})$: the subordinate(s) of holon $\mathsf{h}^{(l)}_{i}\in\{\mathscr{h}^{(l)}_{i}, \bm{\mathscr{h}}^{(l)}_{i}\}$ in the holarchy.
    \item $\text{Nei}(\mathsf{h}^{(l)}_{i})$: the neighbor(s) of holon $\mathsf{h}^{(l)}_{i}\in\{\mathscr{h}^{(l)}_{i}, \bm{\mathscr{h}}^{(l)}_{i}\}$ according to the graph $\mathcal{G}^{(l-1)}_{j}$ s.t. $\bm{\mathscr{h}}^{(l-1)}_{j}\in\text{Sup}(\mathsf{h}^{(l)}_{i})$.
    \item $\mathcal{F}(\bm{\theta})$: the loss function of the system-level model $\bm{\theta}$.
    \item $\mathcal{A}^{(l)}_{i}$: the aggregation function used by holon $\bm{\mathsf{h}}^{(l)}_{i}$.
    \item $[M_1 | M_2]$: the column-wise concatenation of two matrices $M_1$ and $M_2$.
\end{itemize}
Note that based on the holarchical structure of the system and the presumption that the root (cloud) of the system is at level 0, we have: $\bm{\theta}(t)=\bm{\theta}^{(0)}_{1}(t)$ and $\mathcal{F}(\bm{\theta})=\mathcal{F}^{(0)}_{1}(\bm{\theta})$. 


\section{Holonic Learning Model}\label{sec:hlmodel}
This section delves into the details of the proposed distributed holonic learning model. As stated in section \ref{sec:holon} the holons of a holarchy can be seen as reference points, representing different development stages within the system. In other words, each level of the holarchy is self-contained and consists of autonomous entities with specific interactions and problems, depending on the problem's abstraction.
\subsection{Problem Formulation}\label{sec:HL_formulation}
 The presented holonic learning employs a level-based learning model. In the example depicted in Fig. \ref{fig:example_struct}, for instance, the learning problem on the cloud holon is viewed as the process of optimizing model parameters based on the information that is shared by its two subordinate collaborators without worrying about the details of the other levels. With that said, formally, each individual holon locally seeks the solution to the minimization of its empirical risk function $\mathsf{F}^{(l)}_{i}$. For the terminal holon $\mathscr{h}^{(l)}_{i}$, the problem is defined as:
 \begin{equation}
     \min_{\bm{\theta}} \left[\mathsf{F}^{(l)}_{i}(\bm{\theta}):=\frac{1}{|\mathbb{X}^{(l)}_{i}|}\sum_{j=1}^{|\mathbb{X}^{(l)}_{i}|}\mathcal{F}^{(l)}_{i}(\bm{\theta};\bm{\xi}^{(j)})\right]
 \end{equation}
 and for the higher level holon $\bm{\mathscr{h}}^{(l)}_{i}$, we have:
 \begin{equation}
     \min_{\bm{\theta}} \left[\mathsf{F}^{(l)}_{i}(\bm{\theta}):= \sum_{\mathscr{h}^{(l+1)}_{k}\in \text{Sub}(\bm{\mathscr{h}}^{(l)}_{i})} \phi^{(l+1)}_{k} \mathsf{F}^{(l+1)}_{k}(\bm{\theta})\right]
 \end{equation}
 where $\phi^{(l+1)}_{k}$ denote the weight the holon applies to the empirical risk value of the subordinate holon $\mathscr{h}^{(l+1)}_{k}$.
 
 The proposed HoL model tries to find the near-optimal solution for the above-mentioned optimization problem in a sequence of local model updates and aggregation of the feedback it receives from its neighbors, subordinates, superiors. Let $\bm{\theta}_i^{(l)}=\left[\theta_{i,1}^{(l)},\theta_{i,2}^{(l)},\dots,\theta_{i,d}^{(l)}\right]^\top\in\mathbb{R}^d$ specify the model parameter vector of holon $\mathsf{h}^{(l)}_{i}(t)$. For $n$, $b$, and $u$ updates that the holon has received from its neighbors, internal/subordinates, and superiors, respectively, we define the corresponding matrices based the received contributions as follows:
 \begin{equation}
     \bm{\Theta}_{\mathcal{N}_i}^{(l)}(t)=\left[\bm{\theta}_j^{(l)}|\ldots|\bm{\theta}_k^{(l)}\right]\in\mathbb{R}^{d\times n}
 \end{equation}
 and
 \begin{equation}
     \bm{\Theta}_{\mathcal{B}_i}^{(l)}(t)=
     \begin{cases}
         \left[\bm{\theta}_j^{(l+1)}|\ldots|\bm{\theta}_m^{(l+1)}\right]\in\mathbb{R}^{d\times b}&\text{if }\mathsf{h}^{(l)}_{i}(t) \text{ is non-terminal}\\\\
         \left[\bm{\theta}_i^{(l)}\right]\in\mathbb{R}^{d} &\text{if }\mathsf{h}^{(l)}_{i}(t) \text{ is terminal}
     \end{cases}
 \end{equation}
 \begin{equation}
     \bm{\Theta}_{\mathcal{U}_i}^{(l)}(t)=
     \begin{cases}
         \left[\bm{\theta}_j^{(l-1)}|\ldots|\bm{\theta}_o^{(l-1)}\right]\in\mathbb{R}^{d\times u}&\text{if }l>0\\\\
         \text{none} &\text{if }l=0
     \end{cases}
 \end{equation}
 Additionally, each holon uses the column vectors $\bm{w}_{\mathcal{N}_i}^{(l)}(t)\in\mathbb{R}^n$,  $\bm{w}_{\mathcal{B}_i}^{(l)}(t)\in\mathbb{R}^b$, and $\bm{w}_{\mathcal{U}_i}^{(l)}(t)\in\mathbb{R}^u$ to weigh the contributions of its neighbors and subordinates, respectively. With that said, each holon uses its own aggregation function to update the value of its model parameters for the next step as follows:
 \begin{equation}
      \bm{\theta}^{(l)}_{i}(t+1) =\mathcal{A}^{(l)}_{i}\left(\bm{\Theta}_i^{(l)}(t),\bm{W}_i^{(l)}(t)\right)
 \end{equation}
 where $\bm{\Theta}_i^{(l)}(t)$ and $\bm{W}_i^{(l)}(t)$ are referred to as contribution matrix and contribution weights vector, respectively, and are constructed as follows: 
 \begin{equation}
     \bm{\Theta}_i^{(l)}(t):=\left[\bm{\Theta}_{\mathcal{B}_i}^{(l)}(t)|\bm{\Theta}_{\mathcal{N}_i}^{(l)}(t)|\bm{\Theta}_{\mathcal{U}_i}^{(l)}(t)\right]\in\mathbb{R}^{d\times(n+b+u)}
 \end{equation}
 \begin{equation}
     \bm{W}_i^{(l)}(t):=\left[\bm{w}_{\mathcal{B}_i}^{(l)}(t)^\top|\bm{w}_{\mathcal{N}_i}^{(l)}(t)^\top|\bm{w}_{\mathcal{U}_i}^{(l)}(t)^\top\right]^\top\in\mathbb{R}^{(n+b+u)}
 \end{equation}

\begin{figure}
    \centering
    \includegraphics[width=\columnwidth]{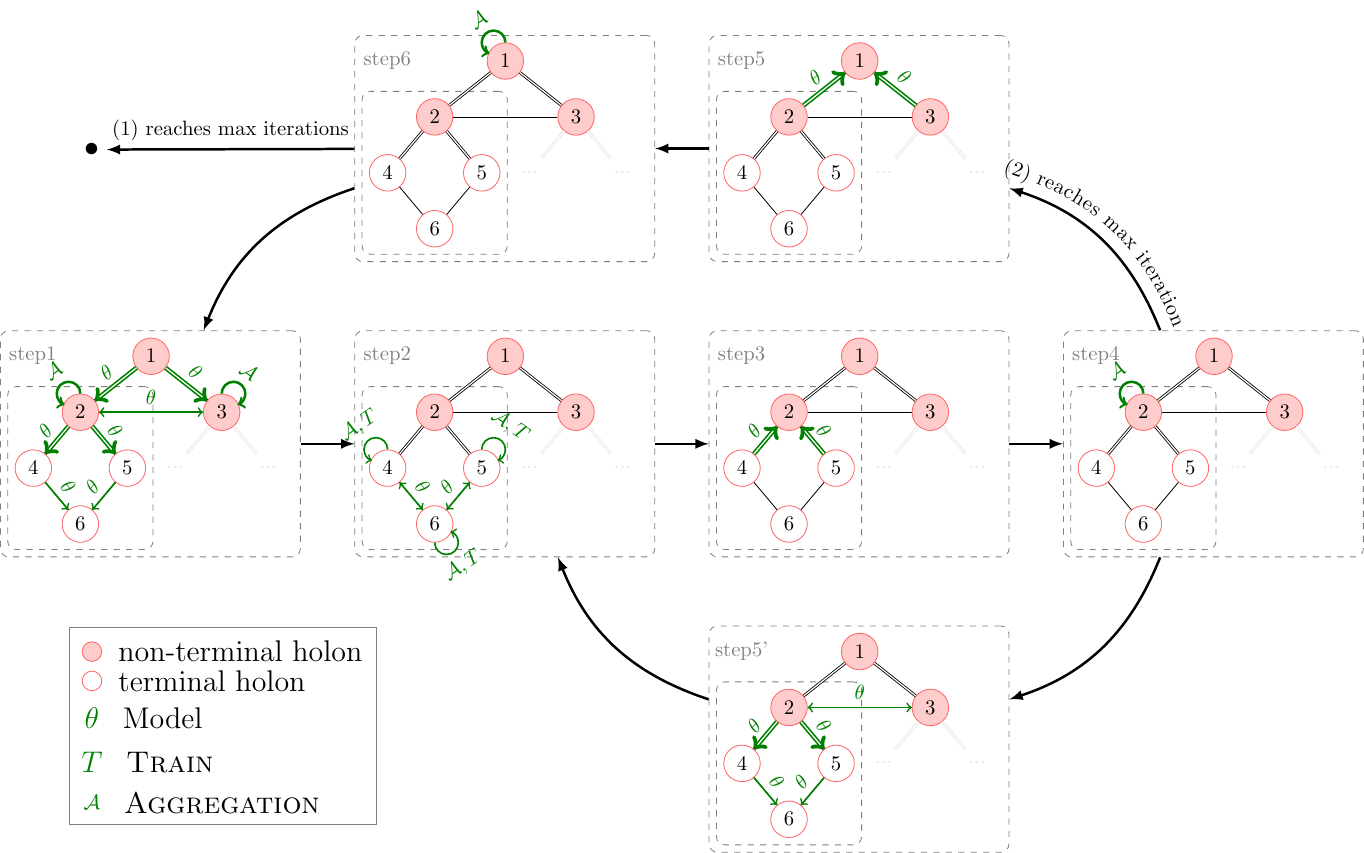}
    \caption{An example demonstrating the flow of communications in Algorithm \ref{alg:HL_model_update}.}
    \label{fig:example_update}
\end{figure}

 The proposed holonic learning employs a criteria-based iterative learning process that dedicates two separate procedures for horizontal and vertical contributions as presented in Algorithm \ref{alg:HL_model_update}. During its lifetime, holon $\mathsf{h}_i^{l}$ uses lines \ref{ln:alg1-listen-start} to \ref{ln:alg1-listen-end} to continuously listen to new messages from the peers, including its subordinates (if non-terminal), neighbors, and superiors, in parallel and properly updates its knowledge matrices and weight vectors accordingly. In case the holon is training in sync with its peers, it also forwards any sent model from the super-holon to the neighbors. Concurrently, the holon follows a set of aggregation, updating, sharing steps until its termination criteria is satisfied. Function \textsc{ReceivedEnoughPeerModels} checks whether it has received expected number of models from its peers and if it has, it runs aggregation on the models and updates its knowledge base and initiates its training process if it is terminal. Next, through the steps between lines \ref{ln:alg1-share-start} and \ref{ln:alg1-share-end}, the holon determines proper peers to share its train results. Choosing the peers to send the model depends on the training stage of the holon, its position in the holarchy, and the internal messaging policy of the super-holon it is a member of. Fig. \ref{fig:example_update} illustrates the steps of the algorithm in a very simple holarchy. It is worth noting that the synchronization of the communications in this example are merely for the sake of clarity and is not the requirement of the proposed model.

\begin{algorithm}
    \SetAlgoLined
    \DontPrintSemicolon
    \SetKw{KwPar}{in parallel}
    \SetKw{KwNBPar}{non-blocking in parallel}
    \SetKw{KwBreak}{break}
    \SetKw{KwOr}{or}
    \SetKw{KwAnd}{and}
    \SetKwProg{myproc}{Procedure}{}{end}
    \SetKwFunction{learn}{\textsc{Learn}}
    \caption{The generic learning procedure run by each holon ${\mathsf{h}}^{(l)}_{i}$ in parallel.}
    \label{alg:HL_model_update}
    \myproc{\learn{}}{
        \While{$C_i^{(l)}=\text{false}$ \KwNBPar}{\label{ln:alg1-listen-start}
            $\bm{\theta}_j, w_j, sender\gets \textsc{Receive}(\bm{\tilde{\theta}})$\;\label{ln:alg1-pass-start}
             $\bm{\Theta}_{\mathcal{U}_i}^{(l)},\bm{w}_{\mathcal{U}_i}^{(l)}, \bm{\Theta}_{\mathcal{B}_i}^{(l)}, \bm{w}_{\mathcal{B}_i}^{(l)}, \bm{\Theta}_{\mathcal{N}_i}^{(l)}, \bm{w}_{\mathcal{N}_i}^{(l)}\gets \textsc{Process}(\bm{\theta}_j, w_j)$\;\label{ln:alg1-pass-end}
             \If{$sender\in \text{Sup}(\mathsf{h}_i^l)$\KwAnd \textsc{Synchronized}}{
                \For{$n \in \text{Nei}(\mathsf{h}_i^l)$ \KwPar}{
                     $\textsc{SendModel}(n,\bm{\theta}_j)$\;
                }
             }
        }\label{ln:alg1-listen-end}
        \While{$C_i^{(l)}=\text{false}$}{
            \eIf{\textsc{ReceivedEnoughPeerModels}$( )$}{
                    $\bm{\theta}^{(l)}_{i}(t)\gets\mathcal{A}^{(l)}_{i}\left(\bm{\Theta}_i^{(l)},\bm{W}_i^{(l)}\right)$\;
                    $\textsc{UpdateKB}\left(\bm{\Theta}_{\mathcal{U}_i}^{(l)}, \bm{w}_{\mathcal{U}_i}^{(l)}, \bm{\Theta}_{\mathcal{B}_i}^{(l)}, \bm{w}_{\mathcal{B}_i}^{(l)}, \bm{\Theta}_{\mathcal{N}_i}^{(l)}, \bm{w}_{\mathcal{N}_i}^{(l)}\right)$\;
            }{
                \textsc{Wait()}\;
            }
            \If{$\mathsf{h}^{(l)}_{i}$ \textbf{is} terminal}{
                $\bm{\theta}_i^{(l)}(t)\gets \textsc{Train}(\bm{\theta}^{(l)}_{i}(t),\bm{\lambda})$\;
                $\bm{\Theta}_{\mathcal{B}_i}^{(l)}\gets \left[\bm{\theta}_i^{(l)}(t)\right]$\label{ln:alg1-term-end}
            }
            $\mathcal{P}_i^{(l)}\gets \textsc{WhoToSendModel}()$\;\label{ln:alg1-share-start}
            \For{$p \in \mathcal{P}_i^{(l)}$ \KwPar}{
                $\textsc{SendModel}(p,\theta_i^{(l)}(t))$\;
            }\label{ln:alg1-share-end}
        }
    }
\end{algorithm}

 The above-mentioned formulation is defined generically in such a way that it supports the autonomy of holons and not enforcing any specific aggregation or communication schemes, it supports the dynamic aspects of the system, including communication graphs, memberships, peer weighting, and other evolving factors. To demonstrate the utilization of the proposed model, we present Holonic Averaging as a homogeneous special embodiment.

 \subsection{Holonic Averaging Learning}
 There is a wide spectrum of learning models that can be designed given the proposed generic HoL framework. This section delves into the details of a specific instance that exemplifies the principles of the presented model and scrutinizes its corresponding algorithms. The presented Holonic Averaging Learning (HoAL) mimics the aggregation mechanism used in the FAVG approach \cite{mcmahan2017communication} and provides a distributed mechanism for a supervised learning problem. Below, we provide the breakdown of the details of HoAL's design components.

\paragraph{Holonic Structure} The holons in HoAL are heterogeneous with respect to their private data and homogeneous in terms of their learning models and aggregation functions. Additionally, we assume that each holon belongs to a unique super-holon. To put it formally, $\forall i\ne j, \forall l,l'\in \mathbb{N}$ we have:
\begin{equation}
    \mathcal{B}_i^{(l)}\cap\mathcal{B}_j^{(l)}=\emptyset,\quad \mathcal{A}_i^{(l)}=\mathcal{A}_j^{(l')},\quad \mathcal{M}_i^{(l)}=\mathcal{M}_j^{(l')}
\end{equation}

 \paragraph{Communication Weights} As seen before, the weights that each holon assigns to the communication with its neighbors, subordinates, and superior allow for more nuanced control over the influence of other holons and can be dynamically defined based on metrics such as reliability, trustworthiness, data quantity, etc. Presuming the locality of the holons belonging the same connected graph $\mathcal{G}_i^{(l)}$ and that they have similar characteristics in terms of performance, data quality, etc., HoAL defines the inter- and intra-holon communication weights across different levels base on the size of the training data. Formally, let $w_{ij}^{(l)}$ denote the weight that holon $\mathsf{h}_i^{(l)}$ uses for a peer $\mathsf{h}_j^{(l')}\in\mathcal{B}_i^{(l)}\cup\mathcal{N}_i^{(l)}\cup\mathcal{U}_i^{(l)}$. We define $w_{ij}^{(l)}=d_j^{(l)}$, where:
 \begin{equation}
     d_j=\begin{cases}
         |\mathbb{X}_j^{(l')}|& \text{if $\mathsf{h}_j^{(l')}$ is terminal}\\
         \sum\limits_{\mathsf{h}_k^{(l'+1)}\in\mathcal{B}_j^{(l')}}{d_k^{(l'+1)}}& \text{if $\mathsf{h}_j^{(l')}$ is non-terminal}
     \end{cases}
 \end{equation}
 Please note that despite its recursive definition, the weight does not need to be calculated recursively each time it is needed. Each non-terminal holon keeps track of its training size whenever it receives updates from the subordinates.
 
 \paragraph{Updates Stopping Criteria} Each holon $\mathsf{h}_i^{(l)}$ in HoAL employs a numerical threshold $\mathsf{k}_i^{(l)}$ to control the number of its local updates. To put it formally, we have:
 \begin{equation}
     C_i^{(l)}:=t_i^{(l)}>\mathsf{k}_i^{(l)}
 \end{equation}
 
 \paragraph{Aggregation} As stated before, HoAL employs weighted averaging in each holon to aggregate the values of model parameters collected in its contribution matrix. Strictly speaking, we have:
 \begin{equation}
     \mathcal{A}^{(l)}_{i}\left(\bm{\Theta}_i^{(l)},\bm{W}_i^{(l)}\right) := \frac{1}{{\bm{W}_i^{(l)}}^\top \bm{1}_{z}}\left(\bm{\Theta}_i^{(l)}\bm{W}_i^{(l)}\right)
 \end{equation}
 where $z$ denotes the size of vector $\bm{W}_i^{(l)}$ at the time of aggregation, and $\bm{1}_{z}$ represents a $z$-dimensional column vector of ones. In this paper, we utilize mini-batch SGD as the optimizer and assume each training iteration equal to a specific number of epochs at the terminal level.
 
 \subsection{Existing models as special cases}
The strength of holonic learning, in its original definition as presented in section \ref{sec:HL_formulation}, is its flexibility to encapsulate an expansive spectrum of distributed learning scenarios. An example that showcases its versatility is when we further investigate the structure of the HAL. While being a specific instantiation of the HoL with some limitations, HoAL still extends the flexibility of several existing distributed approaches, such as federated averaging, hierarchical federated averaging, and peer-to-peer learning to name a few. Figure \ref{fig:other_models} visualizes a high-level example HoAL model accompanied by its particular embodiments to support the aforementioned standard models. Assuming SGD as the optimization method employed by the terminal holons, the depicted instantations can further be discussed as follows:

\begin{figure}
    \centering
    \begin{subfigure}[b]{0.8\columnwidth}
        \centering
        \includegraphics[width=.5\textwidth]{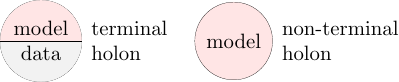}
    \end{subfigure}\\
    \begin{subfigure}[b]{0.4\columnwidth}
        \centering
        \includegraphics[width=\linewidth]{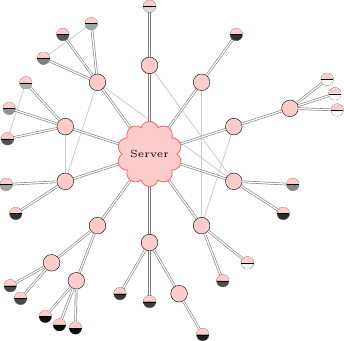}
        \caption{HoAL model}
        \label{fig:hl_model}
    \end{subfigure}\hfill
    \begin{subfigure}[b]{0.4\columnwidth}
        \centering
        \includegraphics[width=\linewidth]{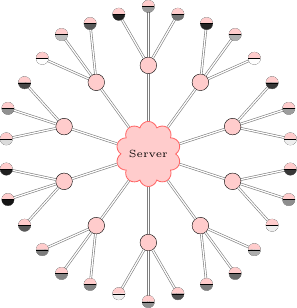}
        \caption{HFL model}
        \label{fig:other_models_hfl}
    \end{subfigure}\\
    \begin{subfigure}[b]{0.4\columnwidth}
        \centering
        \includegraphics[width=\linewidth]{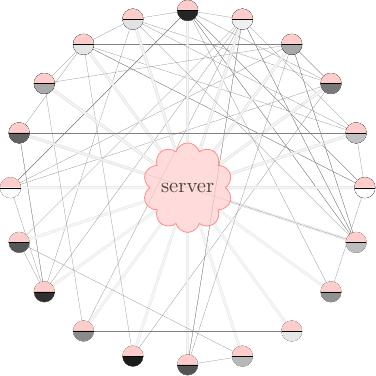}
        \caption{P2P model}
        \label{fig:other_models_p2p}
    \end{subfigure}\hfill
    \begin{subfigure}[b]{0.4\columnwidth}
        \centering
        \includegraphics[width=\linewidth]{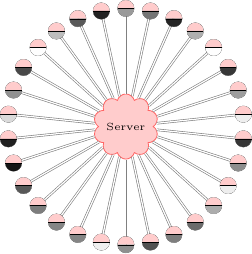}
        \caption{FL model}
        \label{fig:other_models_fl}
    \end{subfigure}
    \caption{The holonic averaging learning model together with different special cases to implement distributed models of federated averaging, hierarchical federated averaging, and peer-to-peer models. Different shades in the lower half of the terminal holons represent the difference in their local data.}
    \label{fig:other_models}
\end{figure}

\paragraph{Federated Averaging}: An FL system can be modeled by a one-level HoAL, where the server is modeled by the root holon $\bm{\mathscr{h}}_1^{(0)}$ and each of the clients with terminal holon ${\mathscr{h}}_i^{(1)}, i=1,\dots,n_1^{(0)}$. Additionally, $\mathcal{G}_1^{(0)}=\overline{K}_n$, where $n$ and $\overline{K}_n$ denote the number of clients and a null graph of size $n$. As depicted in Figure \ref{fig:other_models_fl}, all clients use their own private data to learn the same ML model. At each update carried out by holons $\bm{\mathscr{h}}_1^{(0)}$ and ${\mathscr{h}}_i^{(1)}$ the contribution matrix and weights vectors would respectively be:
\begin{equation}
     \bm{\Theta}_1^{(0)}= \bm{\Theta}_{\mathcal{B}_1}^{(0)}= \left[\bm{\theta}_1^{(1)}|\dots|\bm{\theta}_{n_1^{(0)}}^{(1)}\right]
\end{equation}
\begin{equation}
    \bm{W}_1^{(0)}=\bm{w}_{\mathcal{B}_1}^{(0)}=\left[|\mathbb{X}_1^{(1)}|,\dots, |\mathbb{X}_{n_1^{(0)}}^{(1)}|\right]^\top
\end{equation}
\begin{equation}
     \bm{\Theta}_i^{(1)}= \bm{\theta}_i^{(1)}=\bm{\theta}_i^{(1)}-\eta_i^{(1)}\left[\frac{1}{|\mathbb{X}^{(1)}_{i}|}\sum_{j=1}^{|\mathbb{X}^{(1)}_{i}|}\nabla\mathcal{F}^{(1)}_{i}(\bm{\theta};\bm{\xi}^{(j)})\right]
\end{equation}
\begin{equation}
    \bm{W}_i^{(1)}=\left[|\mathbb{X}_i^{(1)}|\right]^\top
\end{equation}
\paragraph{Peer-to-Peer Learning}: Similar to Fl, a P2P learning model can also be modeled by a one-level HoAL, as depicted in Figure \ref{fig:other_models_p2p}. A key distinction her is that the communication graph connecting all the terminal holons, i.e. $\mathcal{G}_1^{(0)}$ is connected. Additionally, as there is no need for server-side aggregation of the learned models, the root $\bm{\mathscr{h}}_1^{(0)}$, being connected to at least one client, only serves as the initiator of the learning process. Once initiated, the model updates of each holon ${\mathscr{h}}_i^{(1)}$ will be calculated using the following contribution matrix and weight vector configurations:
\begin{equation}
    \begin{split}
        \bm{\Theta}_i^{(1)}= \left[\bm{\theta}_i^{(1)}|\bm{\Theta}_{\mathcal{N}_i}^{(1)}\right]= \Biggl[ \bm{\theta}_i^{(1)}&- \frac{\eta_i^{(1)}}{|\mathbb{X}^{(1)}_{i}|}\sum_{j=1}^{|\mathbb{X}^{(1)}_{i}|}\nabla\mathcal{F}^{(1)}_{i}(\bm{\theta};\bm{\xi}^{(j)})|\\
        \quad\bm{\theta}_e^{(1)}&- \frac{\eta_e^{(1)}}{|\mathbb{X}^{(1)}_{e}|}\sum_{j=1}^{|\mathbb{X}^{(1)}_{e}|}\nabla\mathcal{F}^{(1)}_{e}(\bm{\theta};\bm{\xi}^{(j)})  |\dots|  \\ 
        \bm{\theta}_f^{(1)} &- \frac{\eta_f^{(1)}}{|\mathbb{X}^{(1)}_{f}|}\sum_{j=1}^{|\mathbb{X}^{(1)}_{f}|}\nabla\mathcal{F}^{(1)}_{f}(\bm{\theta};\bm{\xi}^{(j)})\Biggr]
    \end{split}%
\end{equation}
\begin{equation}
    \bm{W}_i^{(1)}=\left[|\mathbb{X}_i^{(1)}|,\bm{w}_{\mathcal{N}_i}^{(1)}\right]^\top=\left[|\mathbb{X}_i^{(1)}|,|\mathbb{X}_e^{(1)}|,\dots,|\mathbb{X}_{f}^{(1)}|\right]^\top
\end{equation}

\paragraph{Hierarchical Federated Learning}: The client-edge-server hierarchical federated learning can be modeled by a two level balanced holarchy, in which the root holon $\bm{\mathscr{h}}_1^{(0)}$ serves as the cloud, the non-terminal holons $\bm{\mathscr{h}}_i^{(1)}, i=1,\dots,n_1^{(0)}$ serves as the edge model-aggregators, and finally the terminal holon ${\mathscr{h}}_j^{(2)}, j=1,\dots,n_i^{(1)}$ represent the clients employed to learn optimize the models using their own data.
Additionally, all the communication graphs of the holons are null, i.e. $\mathcal{G}_1^{(0)}=\overline{K}_{n_1^{(0)}}$ and $\mathcal{G}_i^{(1)}=\overline{K}_{n_i^{(1)}}, i=1,\dots,n_i^{(1)}$. Similar to the above-mentioned federated averaging case, the contribution matrices and the weight vectors would be as follows:

\begin{equation}
    \begin{split}
        \bm{\Theta}_1^{(0)}= \bm{\Theta}_{\mathcal{B}_1}^{(0)}= \left[\bm{\theta}_1^{(1)}|\dots|\bm{\theta}_{n_1^{(0)}}^{(1)}\right]=&\left[\mathcal{A}_1^{(1)}\left(\bm{\Theta}_1^{(1)},\bm{W}_1^{(1)}\right),\dots,\right.\\&\left.\quad\mathcal{A}_{n_1^{(0)}}^{(1)}\left(\bm{\Theta}_{n_1^{(0)}}^{(1)},\bm{W}_{n_1^{(0)}}^{(1)}\right)\right],\quad 
    \end{split}
\end{equation}
\begin{equation}
     \bm{W}_1^{(0)}=\bm{w}_{\mathcal{B}_1}^{(0)}=\left[\sum_{\mathscr{h}_j^{(2)}\in\mathcal{B}_1^{(1)}}|\mathbb{X}_j^{(2)}|,\dots,\sum_{\mathscr{h}_j^{(2)}\in\mathcal{B}_{n_1^{(0)}}^{(1)}}|\mathbb{X}_j^{(2)}|\right]^\top
\end{equation}
\begin{equation}
     \bm{\Theta}_i^{(1)}= \bm{\Theta}_{\mathcal{B}_i}^{(1)}= \left[\bm{\theta}_1^{(2)}|\dots|\bm{\theta}_{n_i^{(1)}}^{(2)}\right]
\end{equation}
\begin{equation}
    \bm{W}_i^{(1)}=\bm{w}_{\mathcal{B}_i}^{(1)}=\left[|\mathbb{X}_1^{(2)}|, \dots,|\mathbb{X}_{n_i^{(1)}}^{(2)}|\right]^\top
\end{equation}
\begin{equation}
     \bm{\Theta}_i^{(2)}= \bm{\theta}_i^{(2)}=\bm{\theta}_i^{(2)}-\eta_i^{(2)}\left[\frac{1}{|\mathbb{X}^{(2)}_{i}|}\sum_{j=1}^{|\mathbb{X}^{(2)}_{i}|}\nabla\mathcal{F}^{(2)}_{i}(\bm{\theta};\bm{\xi}^{(j)})\right]
\end{equation}
\begin{equation}
    \bm{W}_i^{(2)}=\left[|\mathbb{X}_i^{(2)}|\right]^\top
\end{equation}

\section{Experiments}\label{sec:experiments}
This section details our experiments investigating the convergence and performance of HoAL under various design scenarios and data distributions. The HoAL framework was implemented using \verb|pytorch| along with its \verb|multiprocess| and \verb|distributed| libraries. We have utilized training loss and test accuracy as the measure of learning convergence and performance.

\subsection{Settings}
The experimental setup comprised 10 terminal holons (agents) organized into four distinct holarchical structures (see Figure~\ref{fig:exp_holarchs}). These structures were chosen based on their shared characteristic of utilizing a collaboration network among terminal holons, representing various facets of collaborations. For instance, HoAL1P resembles a P2P learning setting, primarily relying on inter-holon communications. In contrast, the remaining three structures differ in the number of holarchical levels while confining collaborations to peers within the holons at each level.

Throughout the experiments, personal data for terminal holons and the collaboration network at the terminal level remained constant. We utilized the standard \verb|MNIST| handwritten digits classification dataset, with all terminal holons equipped with a Convolutional Neural Netwrok Model as in \cite{mcmahan2017communication} initialized with the same values for its 21,840 trainable parameters. Terminal holons used mini-batch Stochastic Gradient Descent (SGD) with a batch size of 32 and an initial learning rate of 0.01. Both terminal and non-terminal holons use separate CPU processes to train/aggregate their models in parallel and employ threads to handle inter-holon communications.

\begin{figure}
    \centering
    \begin{subfigure}[b]{0.31\columnwidth}
        \centering
        \includegraphics[width=\linewidth]{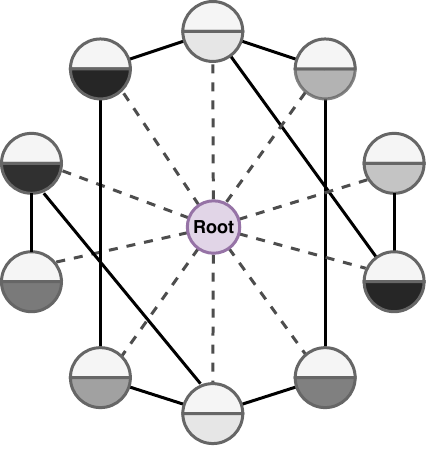}
        \caption{HoAL1P}
        \label{fig:exp_HoAL1P}
    \end{subfigure}\hspace{2em}
    \begin{subfigure}[b]{0.31\columnwidth}
        \centering
        \includegraphics[width=\linewidth]{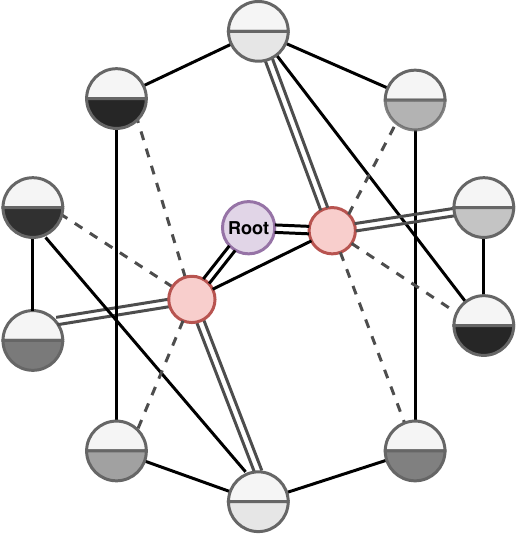}
        \caption{HoAL2L}
        \label{fig:exp_HoAL2L}
    \end{subfigure}\hfill
    \begin{subfigure}[t]{0.15\columnwidth}
        \centering
    \end{subfigure}\\
    \begin{subfigure}[t]{0.33\columnwidth}
        \centering
        \includegraphics[width=\linewidth]{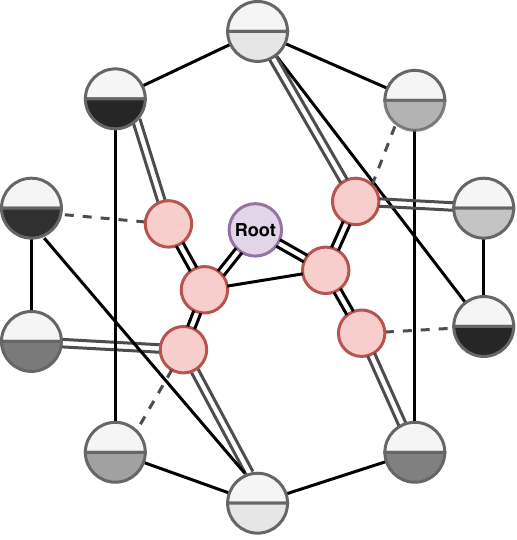}
        \caption{HoAL3L}
        \label{fig:exp_HoAL3L}
    \end{subfigure}~
    \begin{subfigure}[t]{0.33\columnwidth}
        \centering
        \includegraphics[width=\linewidth]{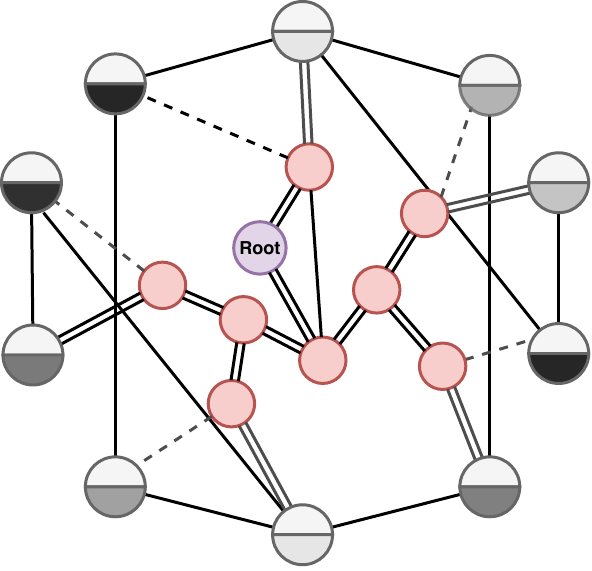}
        \caption{HoAL4L}
        \label{fig:exp_HoAL4L}
    \end{subfigure}~
    \begin{subfigure}[t]{0.25\columnwidth}
        \centering
        \vspace{-14em}\includegraphics[width=.9\linewidth]{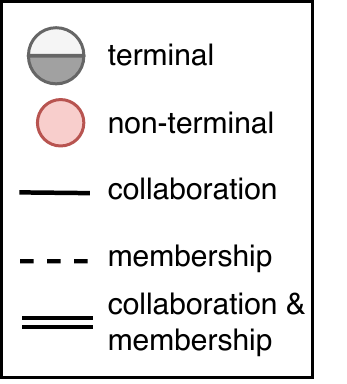}
    \end{subfigure}
    \caption{The different holarchical HoAL structures used for experiments.}
    \label{fig:exp_holarchs}
\end{figure}

Data distribution, a pivotal factor in distributed machine learning, was evaluated in three settings: \verb|IID|, \verb|EqNIID|, and \verb|UEqNIID|. In \verb|EqNIID|, each terminal holon possessed 500 samples from 2 class labels, while in \verb|UEqNIID|, terminals had varying numbers of samples from sorted groups.

Collaboration schemes varied across holarchy levels: each terminal holon underwent 250 rounds of a 5-epoch budget, collaborating with peers and superordinates at the end of each round. Non-terminal holons, on the other hand, had a 2-round local budget for aggregations and guiding subordinates before communicating with their own superordinates. Non-terminal holons budget is reset after each update from their superordinates. To make comparisons fair, we made the duration of experiments be controlled by terminal holons. That is, each terminal holon quits as soon as its budget of 200 rounds is exhausted, and the non-terminal holons quit once all their subordinates finished their job.

\subsection{Results}
\begin{figure}
    \centering
    \begin{subfigure}[b]{.55\columnwidth}
        \centering
        \includegraphics[width=\linewidth]{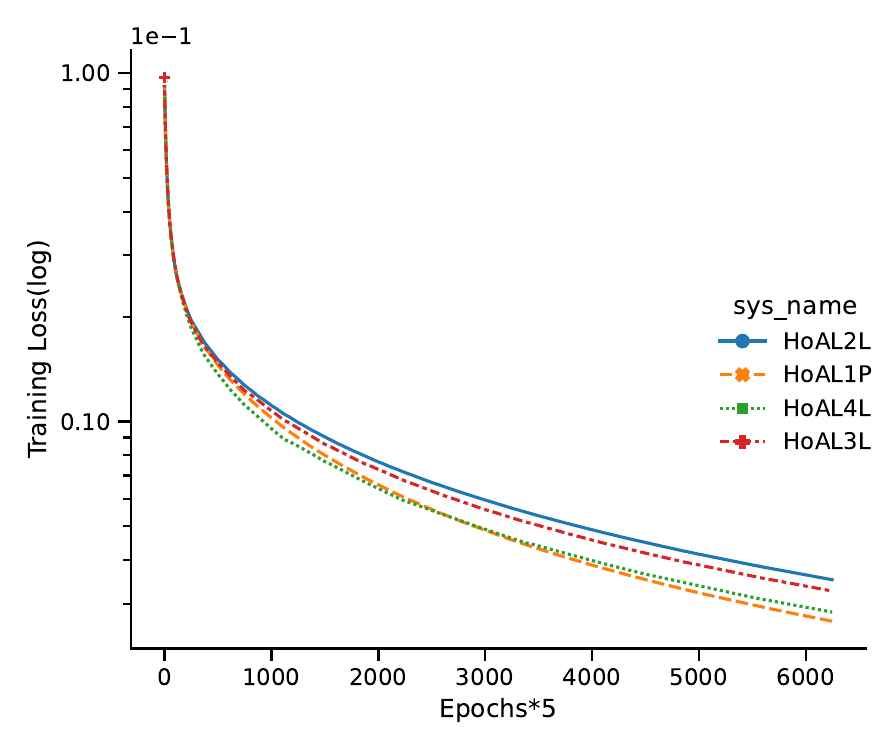}
        \caption{Training Loss}
        \label{fig:res_iid_mnist_train}
    \end{subfigure}\\
    \begin{subfigure}[b]{.55\columnwidth}
        \centering
        \includegraphics[width=\linewidth]{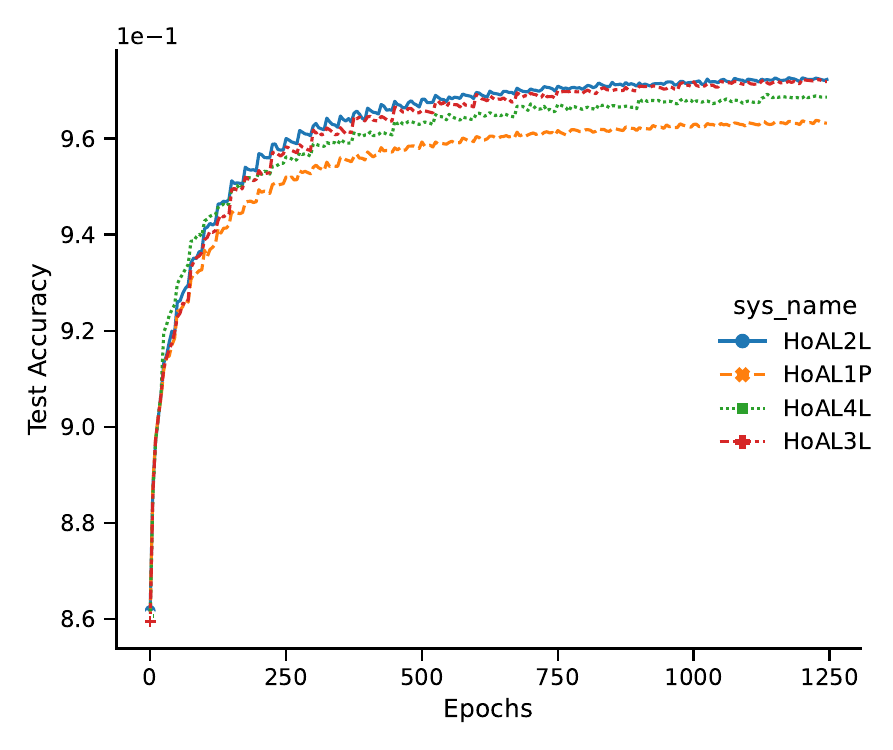}
        \caption{Test Accuracy}
        \label{fig:res_iid_mnist_test}
    \end{subfigure}~
    \caption{The average performance of all terminal holons of different holarchical structures trained on MNIST IID distribution..}
    \label{fig:exp_holarchs}
\end{figure}

In conducting an empirical analysis of HoL's convergence, we observed consistent behavior across various holarchies and data distribution settings. Figures~\ref{fig:res_iid_mnist_train} and \ref{fig:res_iid_mnist_test} present a comparative view of the training loss and test accuracy, respectively, for the experimented holarchies on \verb|IID| data distribution. These metrics are averaged over all terminal holons. As seen, the holarchical structure has yielded relatively improved performance, in both training loss and test accuracy, compared to the plain usage of collaboration network among terminal holons as represented by HoAL1P model.

\begin{figure}
    \centering
    \begin{subfigure}[b]{.55\columnwidth}
        \centering
        \includegraphics[width=\linewidth]{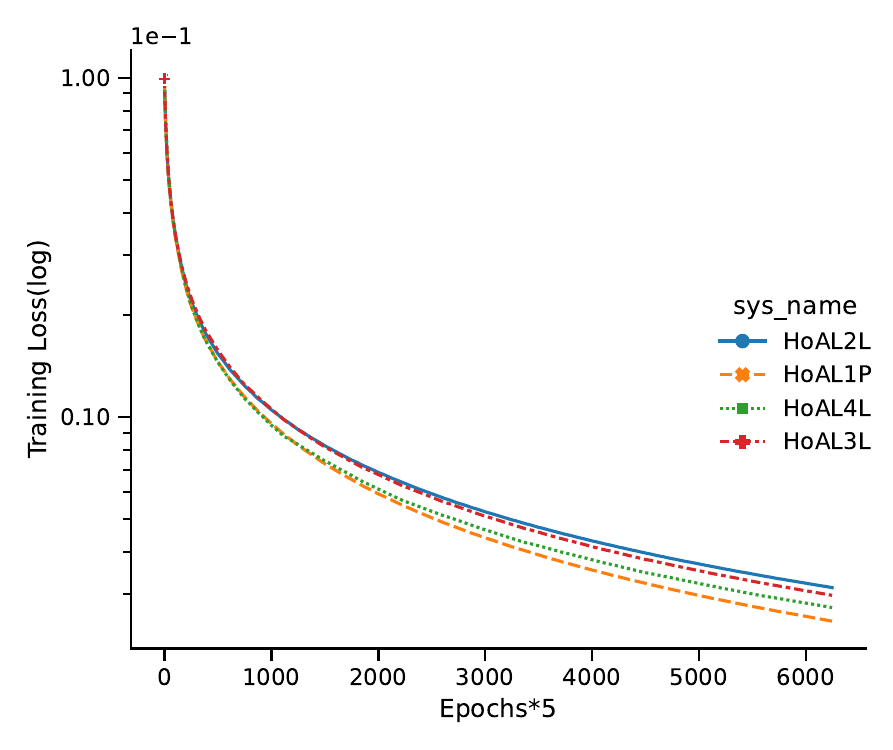}
        \caption{Training Loss}
        \label{fig:res_ueqiid_mnist_train}
    \end{subfigure}\\
    \begin{subfigure}[b]{.55\columnwidth}
        \centering
        \includegraphics[width=\linewidth]{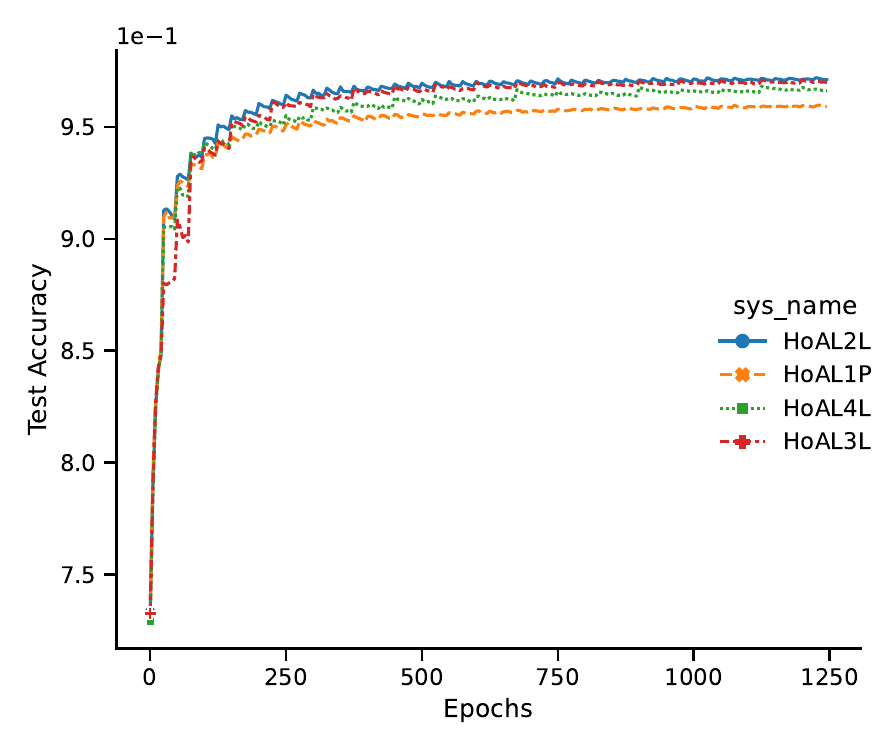}
        \caption{Test Accuracy}
        \label{fig:res_ueqiid_mnist_test}
    \end{subfigure}~
    \caption{The average performance of all terminal holons of different holarchical structures trained on MNIST UENIID distribution..}
    \label{fig:res_ueqiid_mnist}
\end{figure}

This trend persists under the \verb|UEqNIID| data distribution, as demonstrated in Figures~\ref{fig:res_ueqiid_mnist_train} and \ref{fig:res_ueqiid_mnist_test}. Here, the terminal holons exhibit a similar pattern of enhanced performance compared to the collaboration network model, though with comparatively smaller improvements than the \verb|IID| case. The nuances observed in these results indicate the insensitivity of the models to unbalanced data distribution. Such resilience to uneven data distributions further underscores the potential robustness of the HoAL learnings framework.

\begin{figure}
    \centering
    \begin{subfigure}[b]{.55\columnwidth}
        \centering
        \includegraphics[width=\linewidth]{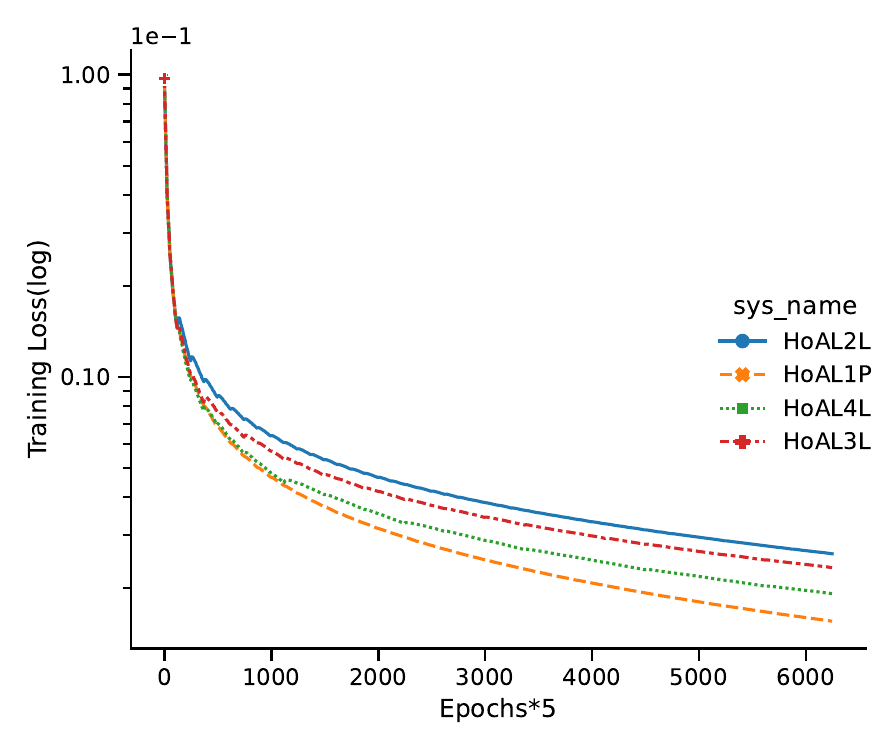}
        \caption{Training Loss}
        \label{fig:res_eqniid_mnist_train}
    \end{subfigure}\\
    \begin{subfigure}[b]{.55\columnwidth}
        \centering
        \includegraphics[width=\linewidth]{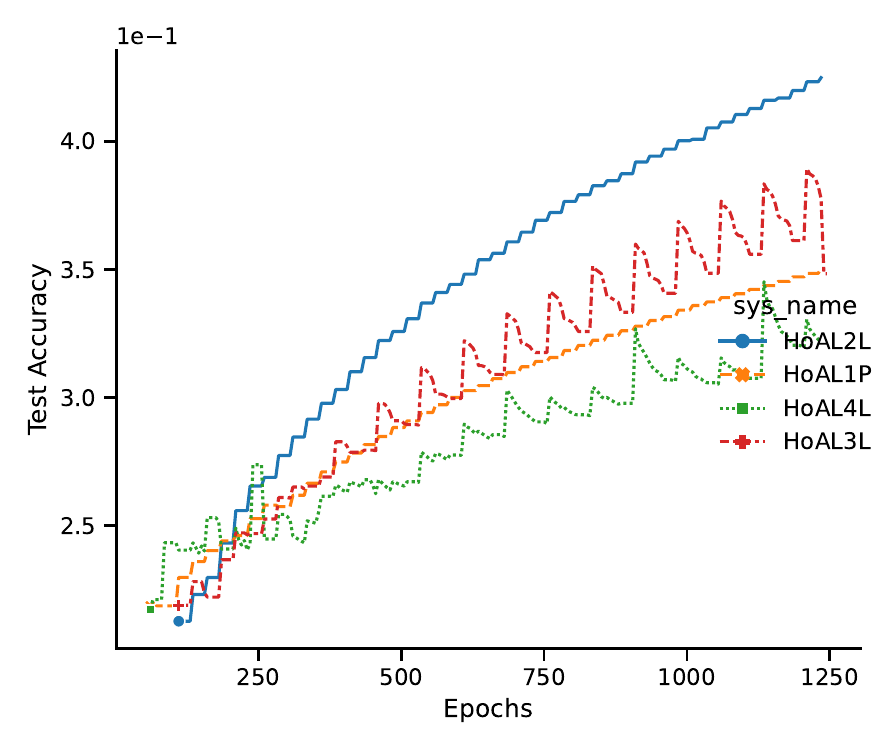}
        \caption{Test Accuracy}
        \label{fig:res_eqniid_mnist_test}
    \end{subfigure}~
    \caption{The average performance of all terminal holons of different holarchical structures trained on MNIST EqNIID distribution..}
    \label{fig:res_eqniid_mnist}
\end{figure}

In the \verb|EqNIID| setting, illustrated in Figures~\ref{fig:res_eqniid_mnist_train} and \ref{fig:res_eqniid_mnist_test}, we persistently observe similar behavior, albeit with more noticeable distinctions in the performance of the examined structures. Unlike the \verb|IID| setting, the introduction of more holarchical levels, as in HoAL4L and HoAL3L cases, has led to a degradation in the generalization capability of terminal holons on average. This observed phenomenon can be primarily attributed to the collaboration delays enforced by extra levels of the holarchy.

While we acknowledge that the presented experimental settings are on a small scale, limiting our ability to fully characterize the extensive potential of the introduced HoL framework, the observed trends and insights serve as a foundational exploration. These initial findings provide a compelling basis for further, larger-scale investigations on the capabilities and robustness of the HoL framework in diverse and complex real-world scenarios.


\section{Conclusion}\label{sec:conclusion}

Inspired by the self-similar holonic paradigm, this paper introduced Holonic Learnings (HoL) as a flexible agent-based distributed machine learning framework. Within HoL, holons are self-contained entities, responsible for managing their individual datasets and model aggregation processes, and its unique characteristic lies in the collaborative synergy across the holarchy, fostering both horizontal and vertical collaborations among holons. This holistic approach not only allows each holon to maintain autonomy over its local data and aggregation strategies but also facilitates the integration of diverse set of policies and learning strategies at different segments and levels of the system. In our exploration of HoL, we presented HoAL, as a homogenous instantiation, featuring a weighted parameter averaging aggregation strategy and showed how it extends the flexibility of existing distributed approaches such as federated learning. The reported investigation of HoAL's capability in various design and data distribution settings revealed its convergence and effectiveness, providing insights into the applicability of HoL.

As we delve into the future of distributed machine learning, Holonic Learnings stands out as a promising model that harmonizes individual agency with collaborative intelligence and, further leveraging the characteristics of holonic multiagent systems, offers vast potentials for the development of scalable, adaptable, privacy-focused, and heterogenous learning environments. This paper used empirical results to scrutinize HoAL's behavior and demonstrated the pivotal role of holarchical design in convergence rate and generalization, even in the presented small-scale settings. In contemplating the next steps, conducting theoretical investigation of HoL's convergence in both convex and non-convex optimization settings and integrating the gained insight into the design of large-scale holarchies are crucial for its real-world applications.



\bibliographystyle{ACM-Reference-Format} 
\bibliography{refs}


\end{document}